\documentstyle[aps,prl,psfig,12pt]{revtex}
\begin{document}
\title{Mean Field Theory for Lyapunov Exponents and Kolmogorov-Sinai
Entropy in Lorentz Lattice  Gases}
\author {M.H.Ernst,
 J.R.Dorfman \thanks{Permanent address: I.P.S.T. and
Department of Physics, University of Maryland, College Park, MD,
20742},  R. Nix, and
D. Jacobs \thanks{Present address: Department of Physics,
Michigan State University, East Lansing, MI, 48824}}
\address{Institute for Theoretical Physics, Utrecht University,
P.O.Box 80006, 3508 TA Utrecht, The Netherlands}
\date{\today}
\maketitle
\begin{abstract}
Cellular automata lattice gases are useful systems for
systematically exploring the connections between non-equilibrium
statistical mechanics and dynamical systems theory. Here the chaotic
properties of a Lorentz lattice gas are studied analytically and
by computer simulations. The escape-rates, Lyapunov exponents and
Kolmogorov-Sinai entropies are estimated for a one-dimensional example
using a mean-field theory, and the results are compared with
simulations for a range of densities and scattering parameters of the
lattice gas. The computer results show a distribution of values for the
dynamical quantities with average values that are in good agreement
with the mean field theory, and consistent with the escape-rate
formalism for the coefficient of diffusion.
\end{abstract}
\pacs{PACS numbers: 05.20.Dd, 05.45.+b,05.60.+w}

The purpose of this letter is to describe a simple model of interest
for non-equilibrium statistical mechanics, the Lorentz lattice gas
(LLG), that allows a detailed
calculation of its chaotic dynamical properties as well as its
transport properties. Here we show that such quantities as Lyapunov
exponents, Kolmogorov-Sinai (KS) entropies, and dynamical partition
functions  can be calculated for LLG's over a full range of densities,
using analytical methods as well as computer simulations both for
closed, periodic and for open systems. The latter case is
of
particular interest since Gaspard and Nicolis \cite{GN} have established  a
relation between the transport
coefficients of hydrodynamics and the Lyapunov exponents and KS
entropies of a fractal set of trapped trajectories in open
systems. Although LLG's are random systems and not of the type usually
accessible by the methods  used widely in dynamical systems
theory, we are able to compute their dynamical properties using
techniques of statistical mechanics. Most of our calculations will focus on the
dynamical partition function which can be used to obtain  all of the  other
dynamical quantities of interest \cite{**,GD}.

A Lorentz lattice gas consists of a moving particle (MP) travelling on
the sites of a $d$-dimensional lattice with unit lattice distance, and
having allowed velocities equal to nearest neighbor lattice vectors.
The dynamical state of the particle at any time $t$ is given by its
position, $\bf r$, and its velocity, $\bf c$. A fixed number of
scatterers $N$ is placed at random on lattice sites, and the density $\rho$
denotes the fraction of occupied sites. We will average over all {\em
quenched} configurations of scatterers.  The dynamics of the MP at
every time step consists of instantaneous collisions with a scatterer,
followed by propagation, in which the MP travels from a site to a nearest
neighbor.  If the MP arrives at a scattering site with velocity $ \bf
c$, it will be transmitted, reflected or deflected with probabilities
$p, q, s$ respectively, normalized as $ p+q+2(d-1) s =1$.  If the MP
arrives at a non-scattering site, its velocity remains unchanged. The
special case of full coverage $(\rho=1)$ is an example of a persistent
random walk, studied by Haus and Kehr \cite{HK}.

The diffusion of the MP in a LLG has been studied in considerable detail, and
the density dependence of the diffusion coefficient is well approximated by a
simple expression, which is exact in the one-dimensional case for all densities
\cite{hvb}.
 Here we address
the chaotic properties of this system.  Although the method described here can
be used for cubic lattices in any dimension, we will consider one-dimensional
systems for simplicity.

We consider a line of lattice points with sites, labeled $ r =
1,2,\ldots ,L$. The sites at $r = 0$ and $r=L+1$ are occupied by absorbers,
so that no particle (re)enters the region $ r = 1,\ldots,L$ from these
sites, and any particle leaving this region is absorbed.
 This construction is of importance for
determining the set of trajectories of the MP that are
trapped forever in the bounded region. The dynamical properties of
this set of trajectories determine the coefficient of diffusion in the
escape-rate formalism.

We begin by considering a quenched configuration of scatterers at density
$\rho$, and define the matrix of transition probabilities $w(x|y)$ to be the
probability for the MP to go from a precollision state  $y =
\{r^\prime,c^\prime\}$ at time $t$ to the precollision state $x =
\{r,c\}$ at time $t+1$.
For our  one-dimensional example, the velocity of the MP $c$ can only take on
the values $\pm 1$. For a quenched configuration of scatterers with
escape the matrix $w(x|y)$ has dimension $2(L-1)$ and is given by
\begin{eqnarray} \label{a1}
& \qquad  w(x|y) = w(r,c|r',c') &\nonumber\\
& = [ a_{1}(r') \delta (c,c') + b_{1}(r') &
\delta (c,-c')]{\delta}(r,r'+c)
\end{eqnarray}
Here $\delta$ denotes a Kronecker delta function;
if there is no scatterer at the point $r'$ then $a_{1}(r')
= 1$ and $b_{1}(r') = 0$; if there is a scatterer, then $a_{1}(r') = p$ and
$b_{1}(r') = q$. In constructing the dynamical partition function we
will need $(w(x|y))^{\beta} \equiv w_\beta (x|y)$, where $\beta$ is an "inverse
 temperature-like" parameter, which might be negative as well. The
random matrix $w_\beta (x|y)$  is also
given by eq (\ref{a1}) with $a_1 (r)$ replaced by $a(r) \equiv a_1^\beta (r)$
and $b_1(r)$ replaced by $b(r) \equiv b_1^\beta(r)$. Further, $w_{\beta}$ is a
periodic matrix \cite{period} with period $2$. This follows
from the dynamics where the particle
will necessarily move alternately from even to odd numbered sites or
{\it vice versa}. This has important consequences for the proper
ergodic decomposition of the ergodic states of this system.

To construct the dynamical partition function we imagine that the MP is
initially placed on the lattice with an initial phase $x_{0} = (r_0, c_0)$
and we ask for the probability $P({\Omega},t;x_{0})$ that
in $t$ time steps the particle follows a trajectory $\Omega= \{ x_0,x_1,
\ldots, x_t\}$ starting at $x_{0}$ and always remaining within the boundaries
of the system. Knowing the hierarchy of $t$--point functions we can
construct the dynamical partition function as \cite{**,GD}
\begin{eqnarray} \label{a2}
Z_{\beta}(L,t;x_{0})& =& \sum_{\Omega}[P(\Omega,t,x_{0})]^{\beta}\nonumber\\
&=& \sum_{x_1,\ldots,x_t}w_{\beta}(x_{t}|x_{t -
1}){\cdots} w_{\beta}(x_{1}|x_{0}).
\end{eqnarray}
With the help of the thermodynamic formalism all dynamical quantities of
interest can be obtained from the {\em topological pressure}
 $\psi_{\beta}(L;x_{0})$, defined by
\begin{equation} \label{a3}
\psi_{\beta}(L;x_{0}) = \lim_{t \to \infty}(1/t)\: \ln
Z_{\beta}(L,t;x_{0}) = \ln \Lambda(\beta,L),
\end{equation}
where $\Lambda(\beta,L)$ is the largest eigenvalue of the random matrix
$w_\beta$. This can be understood by observing that (\ref{a2}) for long times
is essentially the trace of a matrix, {\em i.e.} $(2L)^{-1} {\rm Tr}
(w_\beta)^t$. To obtain the average pressure we have to calculate
the quenched average, $ \langle{\ln Z_\beta}\rangle
$, over all distributions of
scatterers, formally analogous to spin glasses with random interactions
\cite{BG}.
We note that $w_\beta$ is the analog of the transfer matrix in
statistical mechanical calculations of the partition function of Ising-type
lattice models with random interactions. The large--$t$--limit is the analog
of the thermodynamic limit.

The dynamical quantities of interest here are the escape-rate $\gamma(L)$ of
the particle from the lattice, the Lyapunov exponent $\lambda(L)$, and the KS
entropy  $h_{{\rm KS}}(L)$ for the fractal set of trajectories that are forever
trapped on the lattice. According to the thermodynamic formalism these
quantities are given by
\begin{eqnarray} \label{a4}
\gamma (L) &=& -\psi_{1} \nonumber \\
\lambda(L) &=& - \psi_{1}'\nonumber\\
h_{{\rm KS}}(L) &=& \lambda(L) - \gamma(L),
\end{eqnarray}
where the prime denotes a derivative with respect to $\beta$.
For a LLG {\em without} escape $(\gamma =0)$ the Lyapunov exponent is
{\em independent} of the configuration of scatterers, and given by
$\lambda_0 = - \rho p \ln p - \rho q \ln q$ \cite{dej}, and $\lambda_0 =
h_{{\rm KS}}$ (P\'esin's theorem). In a LLG {\em with} escape the exponents
$\gamma(L)$ and $\lambda(L)$ depend on the configurations.

In general it is difficult to evaluate $Z_{\beta}$ or $\Lambda(\beta)$,
except for the special case $\rho = 1$.  Elsewhere we will show that
methods based on the kinetic theory of gases can be used to compute
the various quantities of interest.
 Here we show how simple
mean field and scaling approximations can be used to obtain useful
approximations for $Z_{\beta}$ which compare well with computer
simulations over a wide range of densities and reflection probabilities.
Suppose the
system has a given number of $N$ scatterers distributed in some way over
the $L$ lattice sites. For large $N$ and $L$ the average distance
between scatterers
is $R =  L/N = 1/\rho$. We replace
our random lattice with a regular lattice with scatterers placed a
distance $R$ apart. Thus we can pretend that
we are evaluating the dynamical partition function for a persistent
random walk of a  MP  on an effective lattice of $L/R$ sites and
for an effective time of $t/R$ steps. To
simplify matters further we can
suppose that the initial state $x_{0}$ is located on a scatterer.
Then we obtain the result
\begin{equation} \label{a5}
Z_{\beta}(L,t;x_{0}) \simeq {\cal Z}_\beta(L/R,t/R\,).
\end{equation}
Here ${\cal Z}_\beta$ is the dynamical partition function for a
persistent random walk on a lattice of $L/R ={\cal N}$ sites during  a time
$t/R= {\cal T}$ with absorbing boundaries.

This can be evaluated by noting that
the dynamical partition function for this case is
\begin{equation} \label{a6}
{\cal Z}_\beta({\cal N} ,{\cal T}) \simeq
(1/ 2 {\cal N}) {\rm Tr}(\tilde{w}_\beta)^{{\cal T}},
\end{equation}
where $\tilde{w}_{\beta}$ is the transition matrix for the persistent
random walk, given by
eq (\ref{a1}) with all $a$'s replaced by $a = p^{\beta}$ and all $b$'s by $b =
q^{\beta}$. Since $\tilde{w}_{\beta}$ is a  $2({\cal N}
- 1)$--dimensional matrix with period $2$, its square can be
put in block-diagonal form,
each block corresponding to an invariant, ergodic subspace where
particles on even/odd sites remain on even/odd sites for every {\it
two} time steps. We then look for the eigenstates of each of the two
separate blocks of $\tilde{w}_{\beta}^{2}$ and find that the
largest eigenvalues of both blocks are identical, and denote them by
$\Lambda^2(\beta ,{\cal N})$.  Thus we evaluate
${\cal Z}_{\beta} \simeq \Lambda^{{\cal T}}(\beta, {\cal N})$. For {\em
closed} systems the matrix  $\tilde{w}_\beta$  is
cyclic, and one easily finds $\Lambda(\beta, {\cal N}) = a+b$, from which
 the above result for $\lambda_0$ can be recovered. For
{\em open} systems and large ${\cal N}$ there is a correction of ${\cal O}(1/
{{\cal N}}^2)$, and the topological pressure is found to be,
\begin{equation} \label{a7}
\psi_{\beta}({\cal N} )=
log(a + b) - \frac{a}{2b}\left(\frac{\pi}{{\cal N} + a/b}\right)^2.
\end{equation}
 Our main result is obtained from this equation by scaling ${\cal N}  =  L/R$
and ${\cal T} = t/R$ with the expected free interval between scatterers
 $R = 1/\rho$. We find that for a
LLG with $N$ scatterers, distributed at random over $L$ sites, the
mean field value for the topological pressure is found from (\ref{a3}) and
(\ref{a6}) as
\begin{equation} \label{a8}
\psi_{\beta}(N,L) = \rho \left\{ \ln (a + b) - \frac{a}{2b}
\left(\frac{\pi}{\rho L + a/b}\right)^2 \right\}.
\end{equation}
Using eqs (\ref{a4}) one readily finds
\begin{eqnarray} \label{a9}
\gamma (L) &= & \left( \frac{\rho p}{2q} \right) \left(\frac{\pi}{\rho L + p/q}
\right)^2  \nonumber\\
\lambda(L) &= & \lambda_0 + \left(\frac{\rho p}{2q}\right)
\ln (p/q)\left(\frac{\pi}{\rho L + 2p/q}\right)^2.
\end{eqnarray}
The terms of ${\cal O}(1/L^2)$ are the corrections due to
the absorbing boundary conditions. They have  the  size
dependence expected from the escape-rate formalism \cite{ruek,GN}, which
suggests to write $\gamma(L) \equiv D k^2_0$, where $k_0 = \pi/(L+\rho
p/q)$ is the wave number of the slowest decaying diffusive mode, and
$D= p/(2q \rho)$ is the mean field value of the diffusion coefficient
of the LLG.   This expression is exact for one-dimensional
systems \cite{hvb}.
Combination of $\gamma(L) \equiv D k^2_0$ with the last line of (\ref{a4})
 clearly exhibits the intimate connection
between chaos properties  and  transport properties, as referred to in the
introduction. It has the
same form as the corresponding relation derived  by Gaspard and Nicolis
for the continuous Lorentz gas in a periodic array of scatterers.

\begin{figure}
\centerline{
\psfig{file=cur.eps,height=5cm}}
\caption{\ Probability distribution $P(\ell)$ of reduced Lyapunov exponent
$\ell$ in a LLG with $N=200, L=1000$ and $p=0.2$, taken over $10^7$
configurations. $P(\ell)$ is
a  very broad Gaussian distribution with $\langle{\ell}\rangle=-1.9 \pm 0.3$,
width
$w= 500 |\langle{\ell}\rangle|= 900 $ and kurtosis 0.02.}
\end{figure}

We now discuss the comparison of these results with our computer
simulations.
In order to test to which extent the mean field results capture the
essence of the chaos properties for the LLG, we
 numerically determine the largest eigenvalue of the large
random matrix  $\tilde{w}_{\beta}^{2}$ of linear dimension $2(L-1)$.
In every
quenched distribution of scatterers
the secular determinant of this rather sparse random  matrix is
calculated from  a two step recursion relation.
Its largest root  $\Lambda (\beta, {\cal N})$,
the topological pressure $ \psi_\beta({\cal N})$, and its derivative in
eq (\ref{a9}) are computed numerically. Then we average over  an ensemble
 of different
configurations of $N$ scatterers, placed at random on a lattice of $L$
sites.

For the escape rate the agreement between mean field theory (MFT) and
simulations is excellent, as would be expected, and we first consider the
distribution of escape rates
over the members of the ensemble. It appears to be a {\em very narrow}
Gaussian-type
distribution with a width less than 5\% of the mean. We introduce the reduced
escape rate $g \equiv (\gamma/\rho) (N+p/q)^2$, which is predicted to be
independent of the density $\rho$, the number of scatterers $N= \rho L$ and
system size $L$, and measure it typically over $3 \times 10^6$
configurations of
scatterers. It varies by less than 0.5 \% over the whole density range
$(0.1\leq \rho \leq 0.9)$, and over different system sizes $(N = \rho L
= 50, 100,200, 400)$, and differs by less than 0.5 \% from  the MFT
prediction $\pi^2 (p/2q)$. This also implies that $\langle{ \ln
\Lambda(1)}\rangle$ should be nearly equal to $\ln \langle{\Lambda(1)}\rangle$.
 In fact, they are
found to be equal within statistical errors.
At $N=10$ there are sizable finite size
effects as is to be expected on the basis of (\ref{a9}).

The measured Lyapunov exponents show huge variation over different placements
of scatterers, resulting in a very broad distribution with a width much
larger than the mean.
This is illustrated in Figure 1, which displays the probability
distribution $P(\ell)$ of the reduced Lyapunov exponent, $\ell \equiv (\Delta
\lambda/ \rho) (N+2p/q)^2$, with $ \Delta \lambda = \lambda(L) - \lambda_0$.

The included table compares the MFT results for the reduced
Lyapunov exponent  at different $(N;p)$--values
with $
\langle{\ell}\rangle$, resulting from computer simulations
averaged over  $ 3\times 10^6$  scatterer configurations,
 except at $N=200$ and
$p=0.2$, where $5 \times 10^7$ runs were used, and at $N=200$ and
$p=0.5$, where $2 \times 10^7$ runs were used.  The required CPU time
on a DEC 3000--$\alpha$--machine is typically
$10^{-3} \times N$ days per $10^6$ runs.
\begin{figure}
\hspace{1cm}
\begin{tabular}{|c|c|c|c|c|} \hline
\multicolumn{1}{|c|}{} & \multicolumn{3}{c|}{density  $\rho$}&
\multicolumn{1}{c|}{MFT} \\ \hline
$(N;p)$   & 0.2           & 0.5           & 0.8           & 1\\ \hline
(50;0.2)  & -1.28$^5$ &-1.30$^5$ &-1.40$^2$ & -1.71 \\
(200;0.2) & -1.7$^1$  &-1.4$^1$   & -1.4$^1$  & -1.71\\ \hline
(50;0.5)  & 0.7$^1$   & 0.6$^1$   & 0.5$^1$   & 0 \\
(200;0.5) & 0.3$^3$   & 0.3$^2$   & 0.5$^1$   & 0 \\ \hline
(50;0.8)  & 29.0$^1$  & 28.7$^5$ & 28.4$^5$ & 27.35 \\
(200;0.8) & 27.7$^6$  & 28.1$^4$  & 27.6$^3$  & 27.35  \\ \hline
\end{tabular}
\vspace{5mm}
\caption{Average reduced Lyapunov exponent $\langle{\ell}\rangle$ at different
$(N;p)$--values.
A superscript $a$ denotes a statistical error of $\pm a$ in the last
digit.}
\end{figure}
The variance $w^2 = \langle \ell^2\rangle - \langle{\ell}\rangle^2$ depends
strongly on
$N$ and only weakly on $p$. At $p=0.8$ the ratio $w/\langle{\ell}\rangle$ is
about $6, 20, 100$ for $N=50,200,400$ respectively.  Nevertheless, when
averaging over typically $3 \times 10^6$
runs the mean $\langle{\ell}\rangle$ remains independent of density and system
size, and
is in good agreement (within 2 to 4 \%) with the MFT prediction, $
\langle{\ell}\rangle=
\pi^2(p/2q) \ln(p/q)$. One can also see a small systematic density
variation in the results for $\langle{\ell}\rangle$.
 At small $p$ values ($p \sim 0.2$) the distribution becomes very
broad, as $ \langle{\ell}\rangle$ and $ \langle{\ell}\rangle/w$ decrease by a
factor of 20.
Finite size effects and density dependence of $ \langle{\ell}\rangle$ become
more
pronounced with systematic deviations from MFT up to 7 \%. At $p=0.5$ the
MFT prediction is $ \langle{\ell}\rangle=0$. In an LLG with $N=50$
scatterers, the mean $ \langle{\ell}\rangle$ is non-vanishing ($ w= 300
\langle{\ell}\rangle$),
and shows again a weak, but systematic density dependence. The deviations
from the asymptotic MFT may be ascribed to finite size effects. For larger
systems $(N=200)$ the simulation results are consistent with a vanishing MFT
prediction, but the error bars are very large. Similar broad distributions are
expected to occur in continuous Lorentz gasses with escape \cite{vbd}.

It is important to emphasize two consequences of these very broad
distributions:\\
 (i) The error bars, as listed in the table, have little
bearing on the predictability of the outcome of a single
measurement. For instance, the data used in Figure 1, yields for
$\langle{\ell}\rangle
= -1.9 \pm 0.3$, where the statistical uncertainty $\pm 0.3$ in the average
is calculated over $10^7$ runs. On the other hand, the distribution $P(\ell)$
in
Figure 1  gives only the very weak prediction that the outcome of a single
measurement  $\ell$ will be in the interval $(-1.9 - 900, -1.9 +900)$ with
a 66\% confidence limit.\\
 (ii) According to eqs (\ref{a3}) and (\ref{a4}) the
proper way to obtain mean values of Lyapunov exponents is to calculate
$\langle{\Lambda^\prime (1)/ \Lambda(1)}\rangle$, rather than
$\langle{\Lambda^\prime (1)}\rangle/ \langle{\Lambda(1)}\rangle$. {\em i.e.}
taking the logarithm
and the quenched average
should not be interchanged. Following the latter prescription leads to
values that may be up to factors 6 larger or smaller than
$\langle{\ell}\rangle$,
depending on density and transmission rate.

More details about the derivations of the theoretical
results, the numerical analysis of our sparse matrices with random
elements, and about the effects of  rare configurations
will be published elsewhere. Calculations of topological
entropies, and studies of possible dynamical phase transitions in higher
dimensional LLG's are in progress, as well as extensions of these ideas
to random walks in random environments.

We conclude with the following remarks: \\
 (i) The average values of the dynamical quantities: escape rates, Lyapunov
exponents, and KS entropies (through eq (\ref{a9})), averaged over a large
number of $3 \times 10^6$ to $5 \times 10^7$, are close to  the predictions
of mean field theory
 for all densities, sufficiently large system sizes, and
all model para\-meters descibing the scattering of MP's in LLG's.
The probability distributions of these quantities
 over different placements of scatterers are very broad and of Gaussian
form,
and the mean values do not seem to be affected by rare events.\\
(ii) We have been able to show that a variety of methods from statistical
mechanics can be usefully
applied to determine properties that characterize the chaotic behavior
of non-equilibrium systems. The present results for LLG's are closely
related to  the
calculation presented in a companion paper by van Beijeren and Dorfman
for the continuous Lorentz gas in two dimensions\cite{vbd}. The simplifications
present in LLG models allow
a deeper exploration of systems at high density,
but there are many close connections between the two systems, as well as with
systems with other transport processes taking place.\\
(iii)  A more
systematic approach to the calculations of dynamical quantities for
LLG's can be based on kinetic theory methods. Such a study will lead
to an understanding of the contributions to Lyapunov exponents of the
detailed dynamical
events taking place in the system.

We thank H. van Beijeren, P. Gaspard, C. Beck, H. Bussemaker and R. Brito for
valuable discussions, comments and for help in realizing the computer
simulations.
One of us (D.J.) is financially supported by  the "Stichting Fundamenteel
Onderzoek
der Materie (FOM)", which is sponsored by the "Nederlandse Organisatie voor
Wetenschappelijk Onderzoek (NWO)". JRD thanks FOM for financial
support during a visit in the summer of 1994 to the University of
Utrecht, and acknowledges support from the National Science Foundation
under Grant PHY-93-21312.

\end{document}